# Visualizing "We the People"

## *Bridging the Perception Gap through Pluralistic Data Storytelling*

*Lisa Schirch*
*University of Notre Dame*

*Beth Goldberg*
*Jigsaw*

### Abstract

Traditional visual data storytelling relies on binary graphics that depict two simplified groups in conflict. This can increase political polarization by oversimplifying intra-group disagreements and erasing ambiguity and shared ideas or values. This can inadvertently foster "us versus them" thinking. Intentional, pluralistic design choices for AI-enabled digital platforms can produce visualizations that emphasize nuance, opinion distribution, and intergroup commonalities. To demonstrate this potential, we examine deliberative technologies that map high-dimensional opinion spaces and highlight areas of both consensus and dissensus. The paper highlights the *We the People* deliberation conducted by Jigsaw and the Napolitan Institute in September 2025, which engaged over 2,400 Americans across all 435 congressional districts in an AI-supported, asynchronous dialogue regarding freedom and equality. By utilizing AI to synthesize long-form, text-based participant inputs into interactive "opinion landscapes," the initiative provided an alternative format for pluralistic data storytelling that humanized diverse viewpoints and revealed hidden areas of substantial broad consensus. The paper concludes that shifting from divisive, contrast-heavy visual frameworks to distribution-focused, interactive models represents a highly scalable, low-cost intervention capable of bridging perceptual gaps and cultivating a more resilient, collaborative democratic culture.

### Introduction

"We the People" are the three opening words of the Preamble to the United States Constitution, framing a nation built from a multitude of distinct voices. For generations, maps of the US population on almost any topic have always been complex and colorfully heterogeneous. This is by design, as we are a pluralistic democracy with diverse viewpoints and ways of life. But in the last few decades, maps have become less colorful, more two-tone. Most political visuals, such as electoral maps, polling charts, and partisan graphics, depict a binary US divided into red and blue. These binary data visualizations reinforce "us" and "them" conceptions of American society. They influence us to see each other as opponents rather than as participants in a shared project. They reduce and flatten our complex society. In this visual data storytelling, the idea of “We the People” disappears, replaced by a perpetual “us versus them.”

The visual display of survey results, online discussion platforms, and media outlets can either exacerbate or reduce political polarization. In the world of digital platform UX (User Experience) and software architecture, intentional design choices are the deliberate decisions creators make to influence how users behave, feel, or process information. It affects what users see and what they can do on a platform.

Visualizations on digital platforms and media outlets can not only reify polarized realities but also contribute to the “perception gap,” leading partisans to believe their beliefs are even more divergent than they actually are. On the flip side, data visualizations can prompt more appreciation for pluralism and nuance.

As digital spaces increasingly default to a visual architecture that highlights division, characterizes binary partisan maps, and presents zero-sum polling data, there is a critical need for strategies that prioritize nuance, shared citizen agency, and how individuals navigate and perceive pluralism.

This paper argues that AI-enabled deliberative platform design enables pluralistic data storytelling that can help correct the architecture of division. This type of digital storytelling illustrates how small, individual points create a larger picture or mosaic that represents a more full spectrum of views, where public opinion is not a binary that leads people to sort themselves into "us vs them."

Intentional design choices can bridge the "*perception gap*" and foster a more resilient, collaborative democratic culture. On digital deliberative platforms, these design choices aim to support listening at scale and reflection on other points of view and what is generally known as "civil discourse." These platforms offer participants an opportunity for a structured, safe encounter with difference. These design choices also aim for understanding and collaboration rather than attacking or undermining others (Schirch 2025). The ultimate goal is a more resilient, collaborative democratic culture able to respond to internal and external stresses and challenges with collective problem-solving (Merkel and Lührmann 2021).

## The Visual Architecture of Division: How Chart Design Fuels Polarization

The Pew Research Center study shows that from 1994 to 2017, Americans became significantly more ideologically polarized, with the moderate “middle” shrinking and more people clustering at consistently liberal or conservative positions. The share of Americans with consistently ideological views roughly doubled. At the same time, overlap between Democrats and Republicans nearly disappeared, meaning the typical member of each party now holds views far from the other. At the same time, partisan animosity intensified, with many Americans increasingly viewing the opposing party not just as wrong but as a threat to the nation. Importantly, this polarization is driven most strongly by politically engaged citizens, amplifying the voices of more extreme actors in the public sphere (Pew Res. Cent. 2017).

Polarization is real. But how we visualize it matters. Yale University information design expert Edward Tufte showed that the way we present information can either distort reality or illuminate its complexity. Tufte revolutionized design thinking in his book Envisioning Information by stressing clarity, precision, and integrity in visual communication. Tufte encouraged clear presentation of complexity through a series of intentional design choices that have become best practice across many industries. As digital platforms evolved, his influence became foundational in journalism, civic data, and dashboard design, even as modern interfaces often diverged from his ideals. His work critiques polarizing visuals and guides accurate visualizations of collective life by not oversimplifying (Tufte 1990). Tufte's work is essential for understanding how better digital visualizations can move us beyond reductive “us vs. them” thinking and help reduce polarization.

How media outlets and platforms display data is just as important as the data itself. When news outlets display polling data, they often prioritize "clarity" and "contrast." But in doing so, they may be inadvertently training us to see ourselves as more divided than we actually are.

Traditional "partisan gap" visualizations (such as red vs. blue bar charts) encourage individuals to believe that there are only two points of view represented by two “teams.” This binary can lead community members to take more extreme views to align with their "team." A confluence of factors drives people to embrace a "zero-sum" worldview, in which the other side is seen as a fundamental threat rather than a group with legitimate, albeit different, concerns.

This method of digital storytelling may have significant impacts on US society. Americans consistently overestimate the extremity and hostility of their political opponents. Individuals consistently overestimate the magnitude of disagreement and ideological consistency between political groups, leading to “false polarization.” Distorted beliefs about an opposing party's actual positions lead to significant "perception gaps".

According to the *Perception Gap* report by More in Common, Democrats and Republicans imagine that nearly twice as many people on the other side hold extreme positions as they actually do. For instance, while partisans on average believe that 55% of their opponents hold extreme views, the actual figure is closer to 30%. This distortion leads to a pervasive sense of suspicion and distrust, as each side mistakenly views the other as a fundamental threat to democracy rather than just a group with differing political opinions. This gap is further widened by factors such as high media consumption and higher education among certain groups. The report found that individuals who consume news "most of the time" are nearly three times as likely to be inaccurate in their perceptions as those who only follow the news occasionally. Additionally, highly educated Democrats were found to have a wider Perception Gap, partly due to lower "friendship diversity," where they primarily associate with those who share their political beliefs. This visual and psychological architecture of division fuels a "zero-sum" view of politics, where the more extreme one considers the other side, the more likely they are to attribute negative character traits such as being "hateful" or "brainwashed" to their political opponents (Yudkin et al. 2019).

Colored maps depicting electoral outcomes are not merely passive representations of political landscapes; they shape perceptions of polarization. In their article "Seeing Red (and Blue)," the researchers showed participants two maps that presented the same information. Electoral maps illustrated state-by-state Presidential election results characterized by stark red (Republican) and blue (Democrat) divisions. Other participants viewed proportional maps that showed each party's level of support in shades of purple. Participants viewing stark red/blue electoral maps perceived the nation as more politically divided, stereotyped the political beliefs of residents of various states more, and saw people holding views in the political minority as less agentic and less likely to vote. Those exposed to electoral maps in shades of purple offered a more balanced perspective. Visual representations of electoral data can shape and distort views on polarization (Rutchick et al. 2009).

In their study, "Polarizing Political Polls," researchers found that when people see data broken down by party (Democrat vs. Republican), they tend to shift their own opinions to align with their "team." This team dynamic creates a feedback loop. By highlighting the distance between parties, the "middle ground" looks like no-man's land. Figure 1 illustrates a typical poll with these characteristics. The visual gap between "sides" is empty, suggesting the impossibility of reconciliation or common ground. At the same time, these red-blue graphics amplify in-group conformity. There is psychological pressure to adopt the party position, even on issues where people might choose a more moderate stance (Holder and Bearfield 2024). Framing data through a partisan lens is not just reporting on polarization. It actively encourages it.

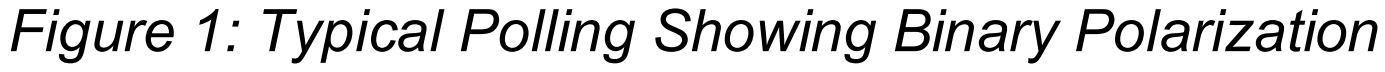

*Figure 1: Typical Polling Showing Binary Polarization*

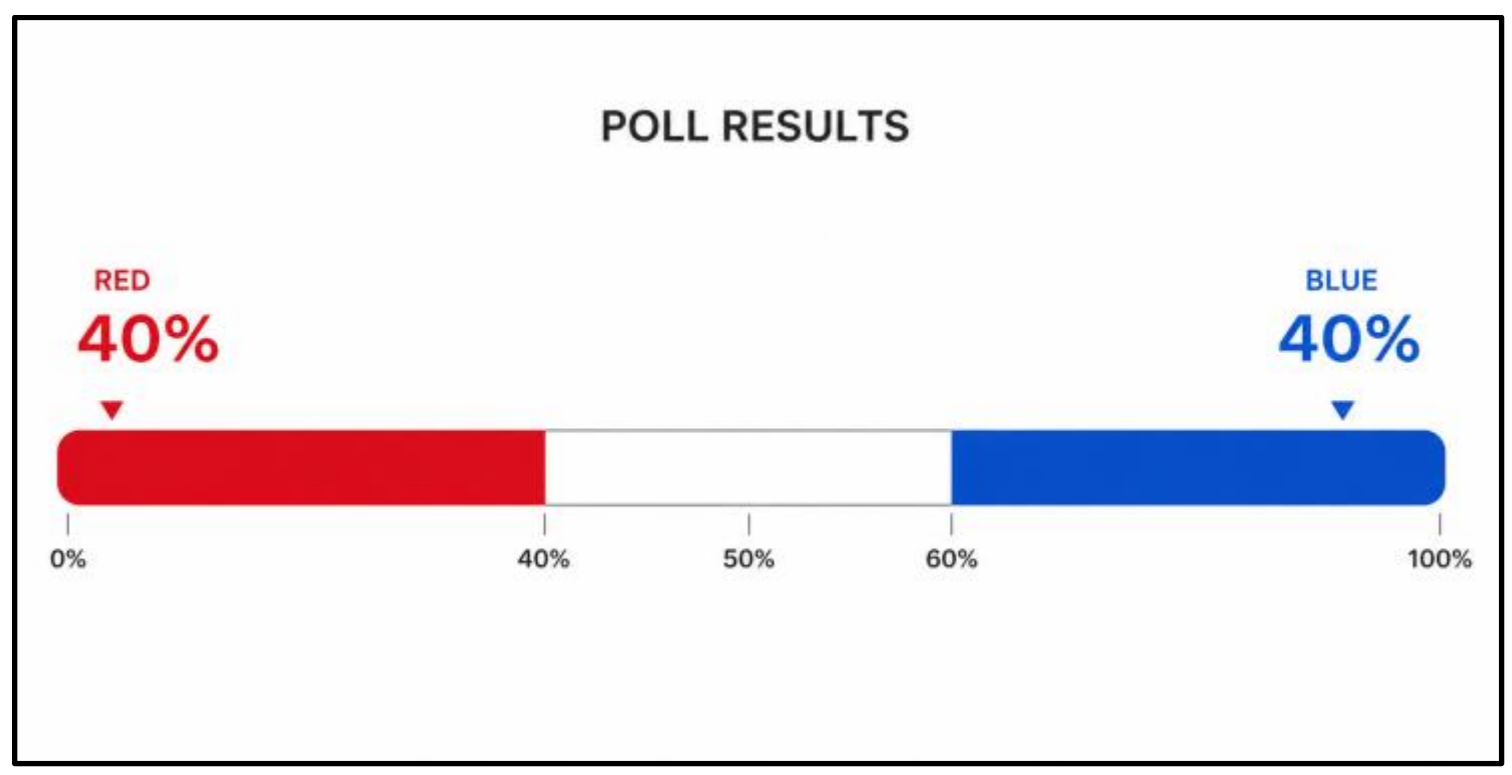

Social psychologists study how people tend to oversimplify and essentialize social groups. Instead of seeing a plurality of differences, humans sort people into different groups and assume wide conformity within those groups. Individuals perceive social categories as distinct and intrinsic rather than overlapping and changing. While they may make exceptions, people assume that members of a group share similar beliefs. This oversimplification can lead to intergroup prejudice and contribute to intergroup conflict (Allport 1954). The stakes of binary red/blue storytelling may be high. Zero-sum thinking is a key driver of political violence and conflicts (Piazza 2023).

**How Americans View Pluralism**

The way communicators display information can invoke change. When information is displayed in a way that highlights a range of perspectives and shared values and beliefs rather than focusing solely on the averages of two conflicting groups, people's perceptions of polarization decrease significantly. The US was born not from consensus, but from sustained engagement across difference. The problem is not disagreement itself, but how we *see* it. The way we structure and present public opinion can actively reduce polarization. Researchers usually present data visually in ways that emphasize difference. This approach leads to abbreviated bar charts that distort research outcomes and failure to recognize or visually portray the vast similarities between people. Researchers find that using different types of visualizations to show the same data can greatly affect how people compare groups. Visuals that clearly show similarities, such as overlapping normal distributions, help people better understand group differences and think more positively about other groups (Hanel et al. 2019).

Ideas for promoting change in attitudes categorize two methods: quick vs careful thinking. The Elaboration Likelihood Model says attitudes change in two ways: a central route, involving careful thought, and a peripheral route, involving quick reactions. Elaborations are thoughts that arise during message processing and can change attitudes, such as thinking about what a message means (Petty and Cacioppo 1986). For example, seeing a graphic depicting red and blue states might lead to a quick thought about polarization in the US. Without interest or ability, people might use the peripheral route, judging the message based on simple cues, such as a graphic depicting a map divided into two colors.

Researchers find that people change their attitudes when they think carefully about an image or chart and connect it to what they know. A simple chart with two colors can lead people to judge quickly without thinking deeply, thereby preventing long-term attitude change. In topics where people have strong opinions, it's important to avoid signs that prompt rejection of the message. Persuasion depends on how well they understand the message (Markant et al. 2022). Other scholars find that the charts with graphic information may be less effective than tables with textual information if people have preconceived beliefs (Pandey et al. 2014).

Recent results from a large-scale study across social media clearly show that how platforms display content is a design choice. Platform and product designers *can* reduce polarization by choosing algorithms that rank or order content (Stray et al. 2026). The research investigates the potential for redesigning social media algorithms to emphasize "prosocial" outcomes, such as decreasing polarization and enhancing well-being, while maintaining user engagement levels. Researchers conducted a real-world, controlled experiment by employing a custom browser extension to rearrange feeds on Facebook, X, and Reddit, testing various ranking strategies, such as "bridging-based ranking," to reduce toxicity. The main conclusion from their recent findings is that it is feasible to significantly reduce polarization and conflict without affecting the time users spend on these platforms.

Researchers find that shifting to more effective visualization formats represents a highly scalable, low-cost intervention to reduce political misperceptions (Tartaglione and de-Wit 2025). A study on digital platforms that display opposing viewpoints side-by-side found that repeated exposure to multiple perspectives significantly increases open-mindedness and reduces extreme positions. Researchers studied a platform called *The Perspective*, which presents two sides of an issue side-by-side. In an experiment with 400 participants, people were exposed to varying amounts of balanced, opposing viewpoints, ranging from a single article to repeated exposure over multiple days. Participants who read multiple articles presenting opposing views became significantly more open-minded than those who saw only one or none. As a result of this treatment, polarization decreased, especially compared to control groups who had no exposure to different points of view. Groups exposed to multiple perspectives showed less extreme ("hardliner") positions and narrower ideological gaps than the control group. The researchers also found that simple nudges or prompts urging participants to be open-minded were less effective than actual repeated exposure to diverse views. Rather than pushing people further apart, thoughtfully designed interfaces can help users recognize complexity and reconsider their assumptions (Einav et al. 2022).

It is striking that the effect of depolarization did not depend on changing people's beliefs. It depended on changing what they *saw*. The visual presentation of different points of view offered a structured encounter with difference. In other words, when people are given a more complete side-by-side view of different perspectives, their views become less polarized. Visualizations that emphasize the distribution and overlap of opinions can reduce perceived polarization and help community members form more accurate estimates of inter-party agreement. When researchers tested 25 interventions head-to-head to decrease US partisan animosity, they found that visual evidence correcting misperceptions of political divides (closing the perception gap) or highlighting common identities shared with rival partisans not only informed more accurate understandings of the other but reduced polarization and improved democratic attitudes (Voelkel et al. 2024).

There is mounting evidence that oversimplifying information is polarizing. In the next section, we explore how illustrating common ground or even just a plurality of opinions might be depolarizing. We can ask, "When we see that many Democrats and Republicans actually share similar views, does the 'us vs. them' narrative start to crumble?" If so, one scalable depolarization strategy might be to provide better data visualizations of cross-party attitudes on key issues. Deliberative technologies could make this strategy possible by surfacing public input on key issues at unprecedented scale, speed, and cost-effectiveness.

**Pluralistic Data Storytelling with Deliberative Technologies**

Digital tools called “deliberative technologies” offer more nuanced data presentation and a chance for structured encounters with difference. Deliberative technologies enable l*arge-scale exchange of views* among the public through iterative discussion, enabling participants to evolve in their understanding (Schirch 2024). These platforms are helping to reimagine a more functional digital public square (Goldberg et al. 2024). Some deliberative technologies generate visualizations of both areas of common ground or shared interests and distinctions or differences. These affordances can help community members transcend traditional "us versus them" narratives.

For example, a platform called Pol.is offers a space for large-scale, structured online dialogue. It uses machine learning to surface areas of consensus within polarized populations. Unlike traditional polls, Pol.is invites participants to submit and vote on short statements. A "bridging algorithm" clusters users based on shared opinions and identifies points of convergence across divides (Ovadya and Thorburn 2023). This approach shifts the focus from adversarial argument to collective sensemaking. Even when disagreement is deep, there are often overlapping values or conditional agreements that can serve as entry points for dialogue and policy innovation.

Pol.is offers three different types of visualization of a digital deliberation. First, each participant's statement is a dot on a spectrum indicating whether participants reached consensus or viewed it as divisive. Second, Pol.is’ machine learning groups similar ideas together in high-dimensional opinion groups (Small et al. 2021). Participants can see how their views compare with others in an autogenerated opinion map on Pol.is as illustrated in Figure X. Third, Pol.is also offers more traditional digital visualization in chart form. The chart in Figure X shows where there is majority agreement. Unshown are charts on the opinion groups and areas of uncertainty.

*Figure 2: Pol.is Spectrum Visualization*

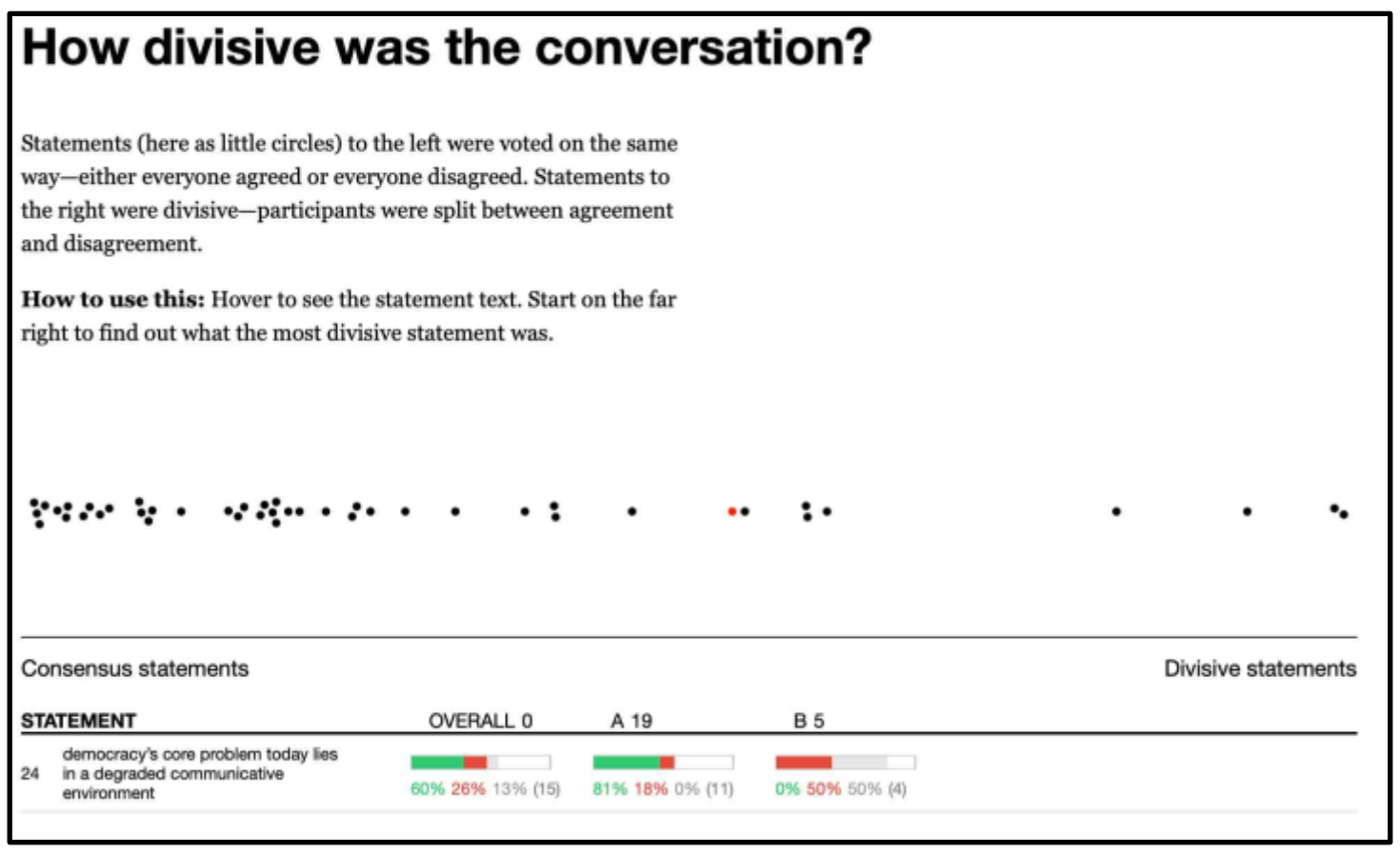

*Figure 3: Pol.is Opinion Map*

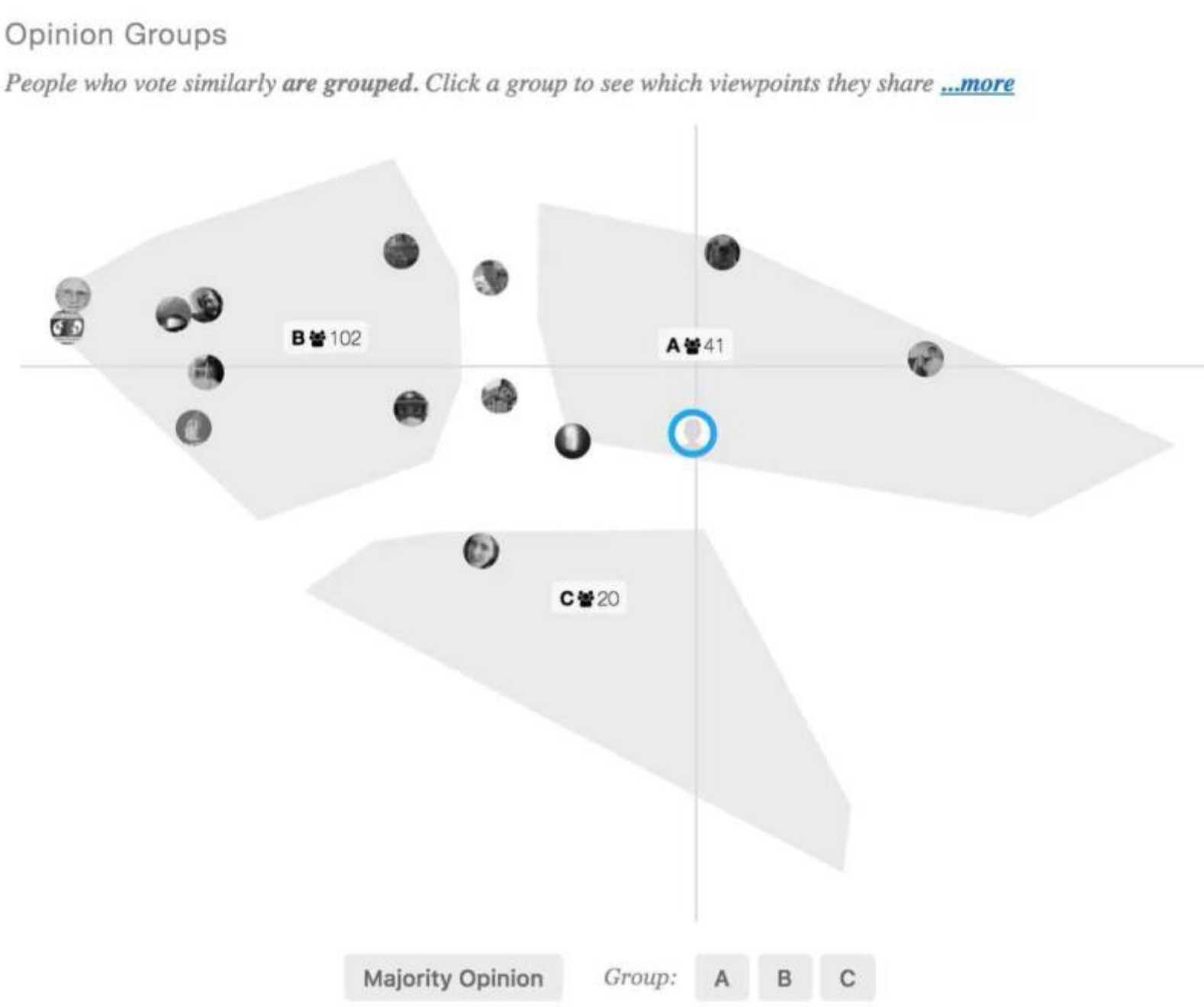

*Figure 4: Pol.is Visualization of Areas of Majority Agreement*

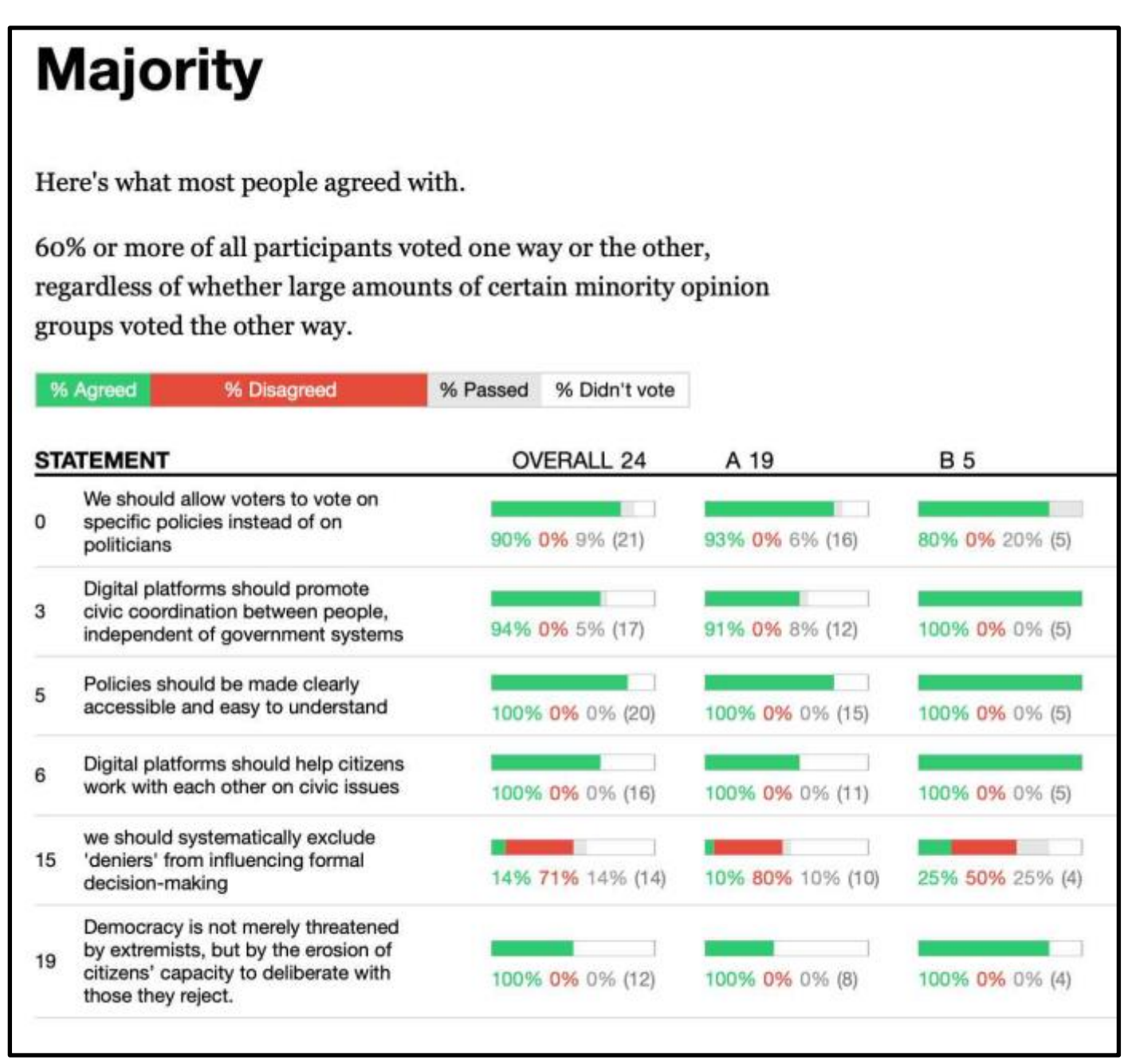

## Majority

Here's what most people agreed with.

60% or more of all participants voted one way or the other, regardless of whether large amounts of certain minority opinion groups voted the other way.

% Agreed | % Disagreed | % Passed | % Didn't vote

| | STATEMENT | OVERALL 24 | A 19 | B 5 |
|---|---|---|---|---|
| 0 | We should allow voters to vote on specific policies instead of on politicians | 90% 0% 9% (21) | 93% 0% 6% (16) | 80% 0% 20% (5) |
| 3 | Digital platforms should promote civic coordination between people, independent of government systems | 94% 0% 5% (17) | 91% 0% 8% (12) | 100% 0% 0% (5) |
| 5 | Policies should be made clearly accessible and easy to understand | 100% 0% 0% (20) | 100% 0% 0% (15) | 100% 0% 0% (5) |
| 6 | Digital platforms should help citizens work with each other on civic issues | 100% 0% 0% (16) | 100% 0% 0% (11) | 100% 0% 0% (5) |
| 15 | we should systematically exclude 'deniers' from influencing formal decision-making | 14% 71% 14% (14) | 10% 80% 10% (10) | 25% 50% 25% (4) |
| 19 | Democracy is not merely threatened by extremists, but by the erosion of citizens' capacity to deliberate with those they reject. | 100% 0% 0% (12) | 100% 0% 0% (8) | 100% 0% 0% (4) |

Similarly, platforms like Consider.it move away from inflammatory direct debate toward "reflective public thought." Consider.it encourages users to create balanced pro/con lists and acknowledge the trade-offs in their own positions. In doing this, participants practice active listening and a better understanding of opposing perspectives. Consider.it provides a spectrum visualization that shows where participants stand on an issue. Rather than asking agree/disagree questions, Consider.it offers participants a slider that allows them to express nuance in their responses (Kriplean et al. 2012).

*Figure 5: Consider.it Spectrum Visualization*

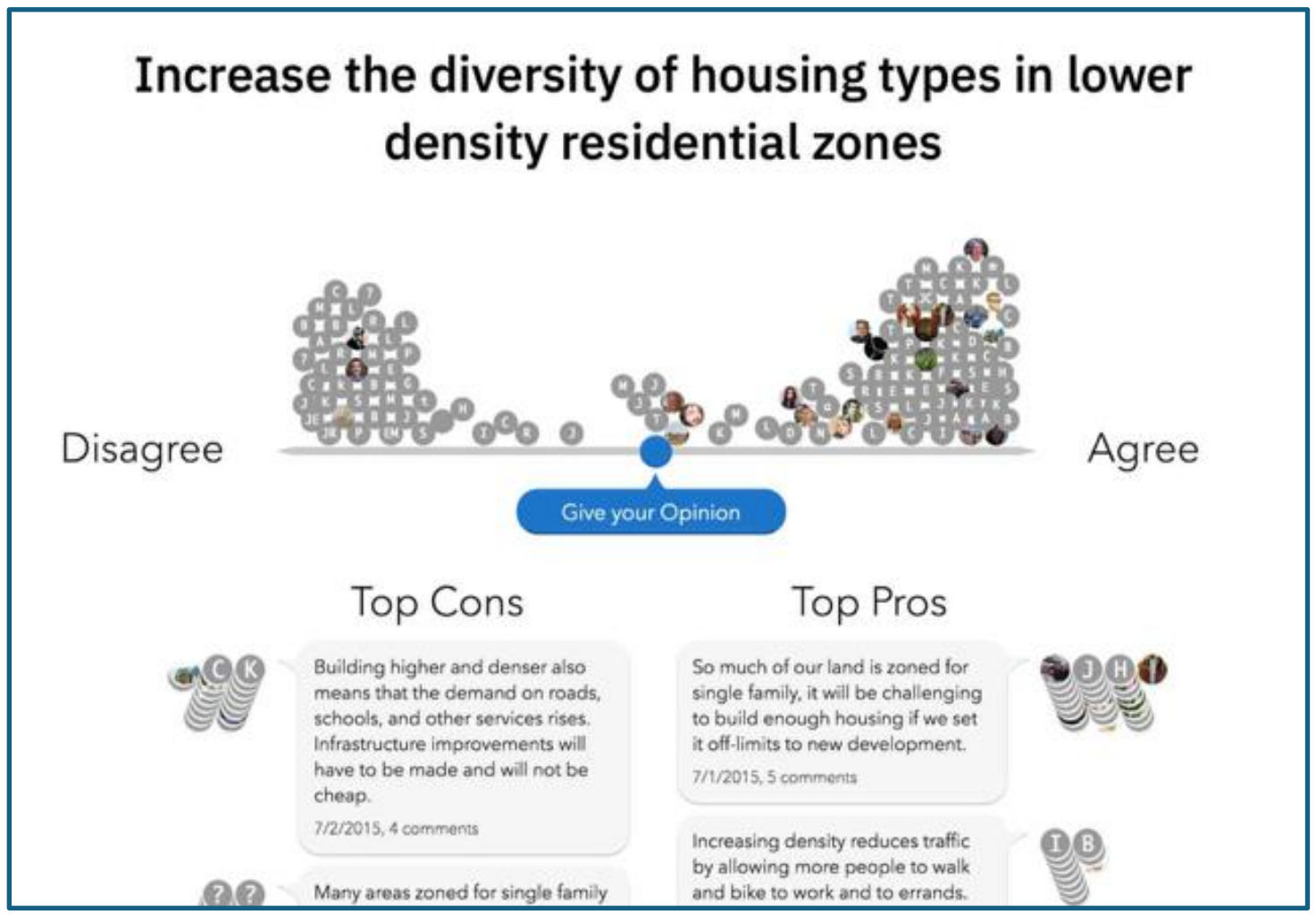

**Jigsaw and Napolitan Institute's *We the People* Deliberation**

The "We the People" initiative by Jigsaw and the Napolitan Institute offers another approach to pluralistic data storytelling. They convened large-scale digital conversations in September 2025 with over 2,400 Americans from all 435 congressional districts to discuss one question: “What do freedom and equality mean to you?”(Jigsaw 2025).

The nationwide conversation is structured as a three-round process: individual expression, collective reflection, and conversation validation. In the first phase, participants are asked a series of questions related to freedom and equality. In contrast to conventional surveys, where respondents are restricted to a predetermined set of choices, participants were able to respond to questions in detail in open text. Then they were prompted dynamically by AI-generated tailored follow-up questions to elaborate on their statement and personal experiences. This provided a much richer, more nuanced dataset than was previously possible using multiple choice surveys, in-person focus groups where fewer people get to talk at length, or other deliberative technologies that solicit single short comments.

The second phase uses AI to summarize thousands of long-form responses into an interactive "opinion landscape," where participants can explore and respond to fellow Americans' views. Finally, the AI generates declarative statements that reflect the group's broad consensus, which participants rate to gauge how well the AI captures their views.

The interactive report on the conversation includes a variety of digital visualization tools that highlight the diverse participant pool’s common values, while holding a wide plurality of views. The interactive report begins by illustrating the opinion landscape of the full conversation had by the 2,400 participants, broken into four major topics, illustrated in Figure 6. Each box represents topics and sub-topics, as summarized by AI and validated with human review. Readers can explore all of the opinions and individual quotes.

*Figure 6: We the People Pluralistic Data Storytelling via Opinion Maps*

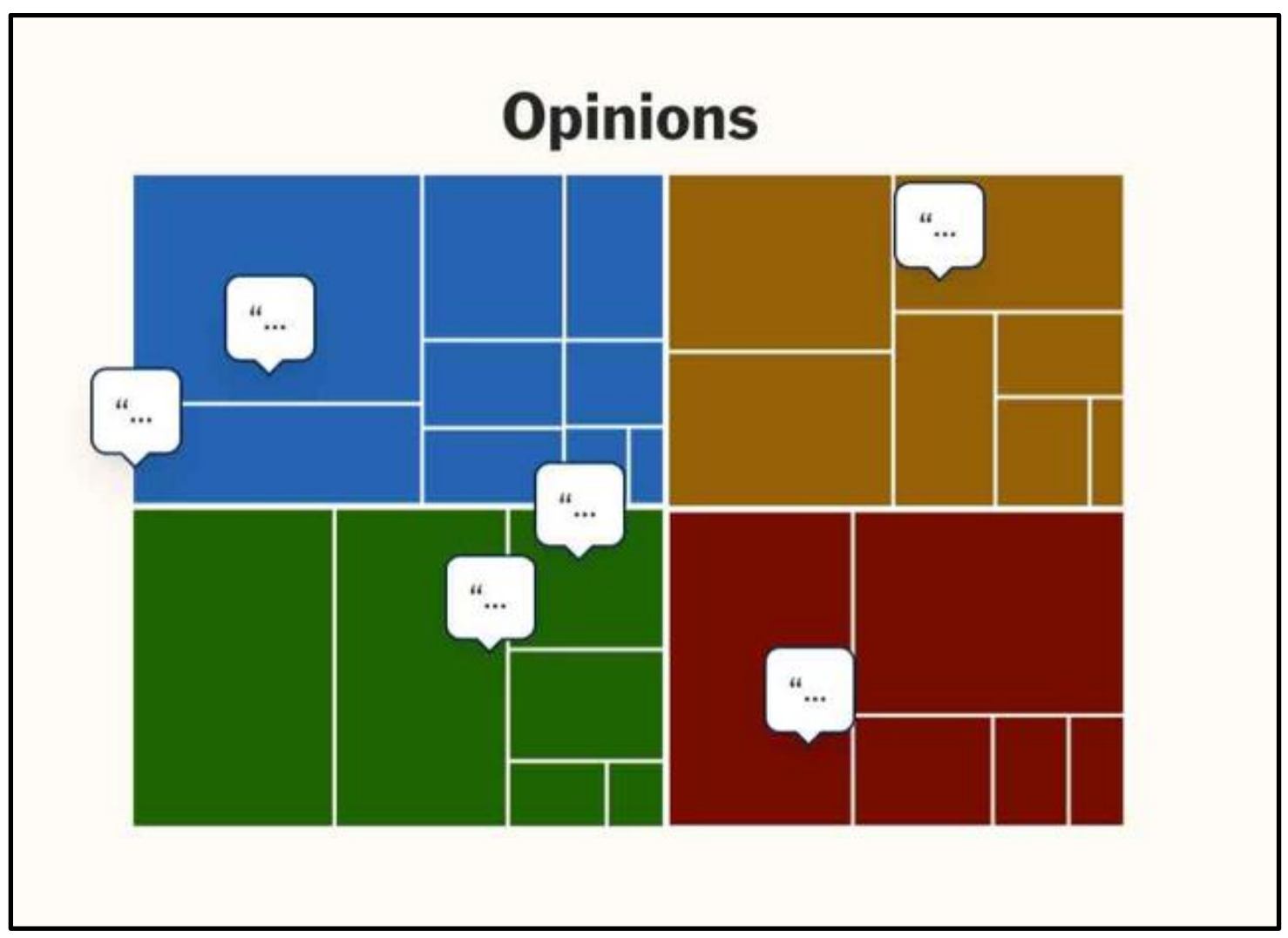

*Figure 7: We the People Opinion Map A*

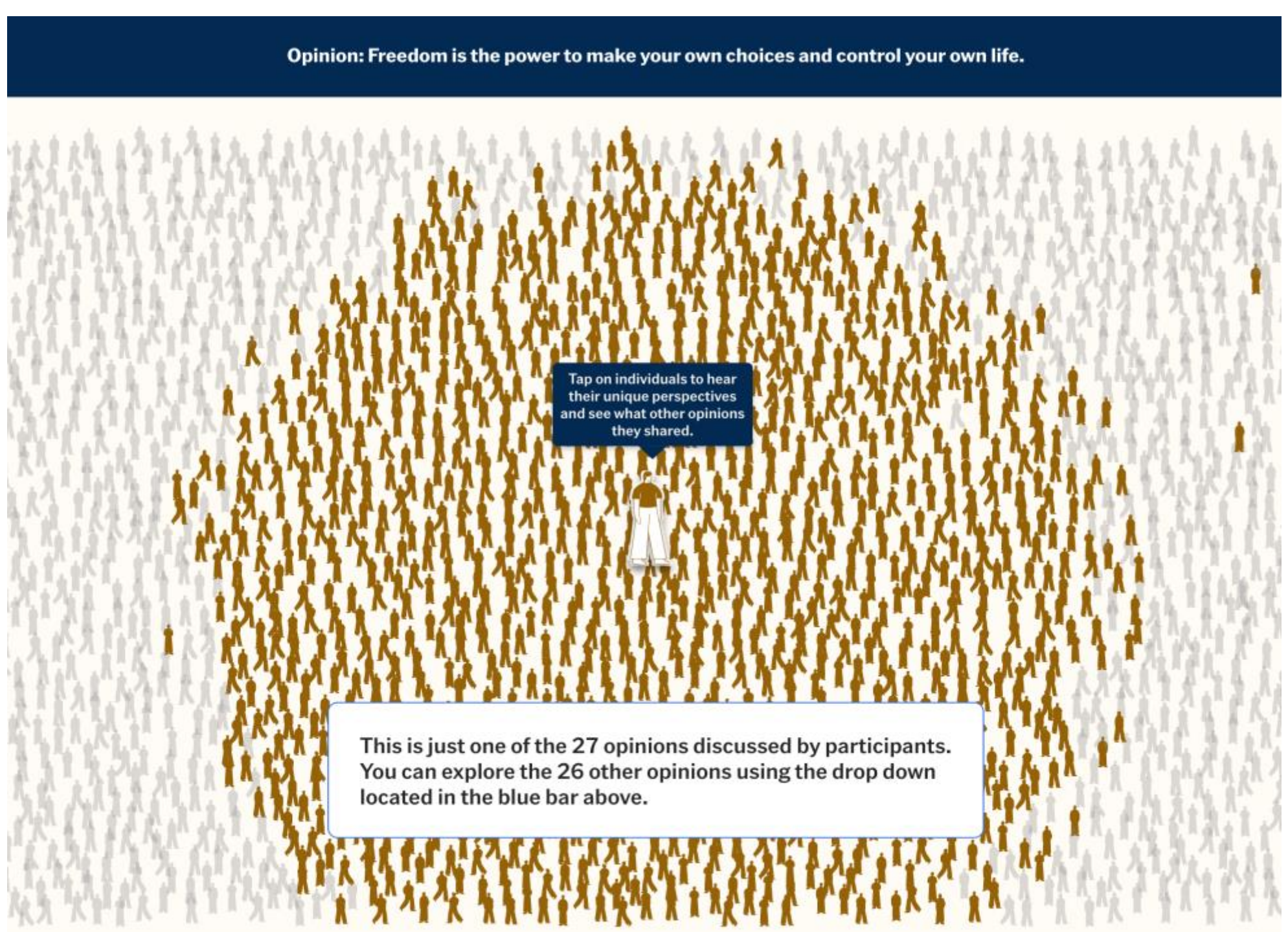

The interactive report then invites viewers through the three phases of the participant experience from the perspective of a single participant. By data storytelling from a single viewpoint, this visualization journey is intended to humanize diverse perspectives. The viewer is invited to click on an icon of an individual participant representing an American who expressed thoughts relevant to the opinion at the top of the screen. By making the data visualization interactive and enabling the viewer to select any and all participants' quotes, the data storytelling intends to maintain the agency of the viewer and avoid bias by only including cherrypicked voices. These intentional design choices ensure the full plurality of voices is available to the viewer as they journey through the data.

Once the viewer taps on a single individual in the crowd and scrolls,  the visualization allows viewers to "listen" to someone in the crowd at random. These quotes were often very personal and evocative, illustrating how political values are experienced in a breadth of ways that are not all polarizing.

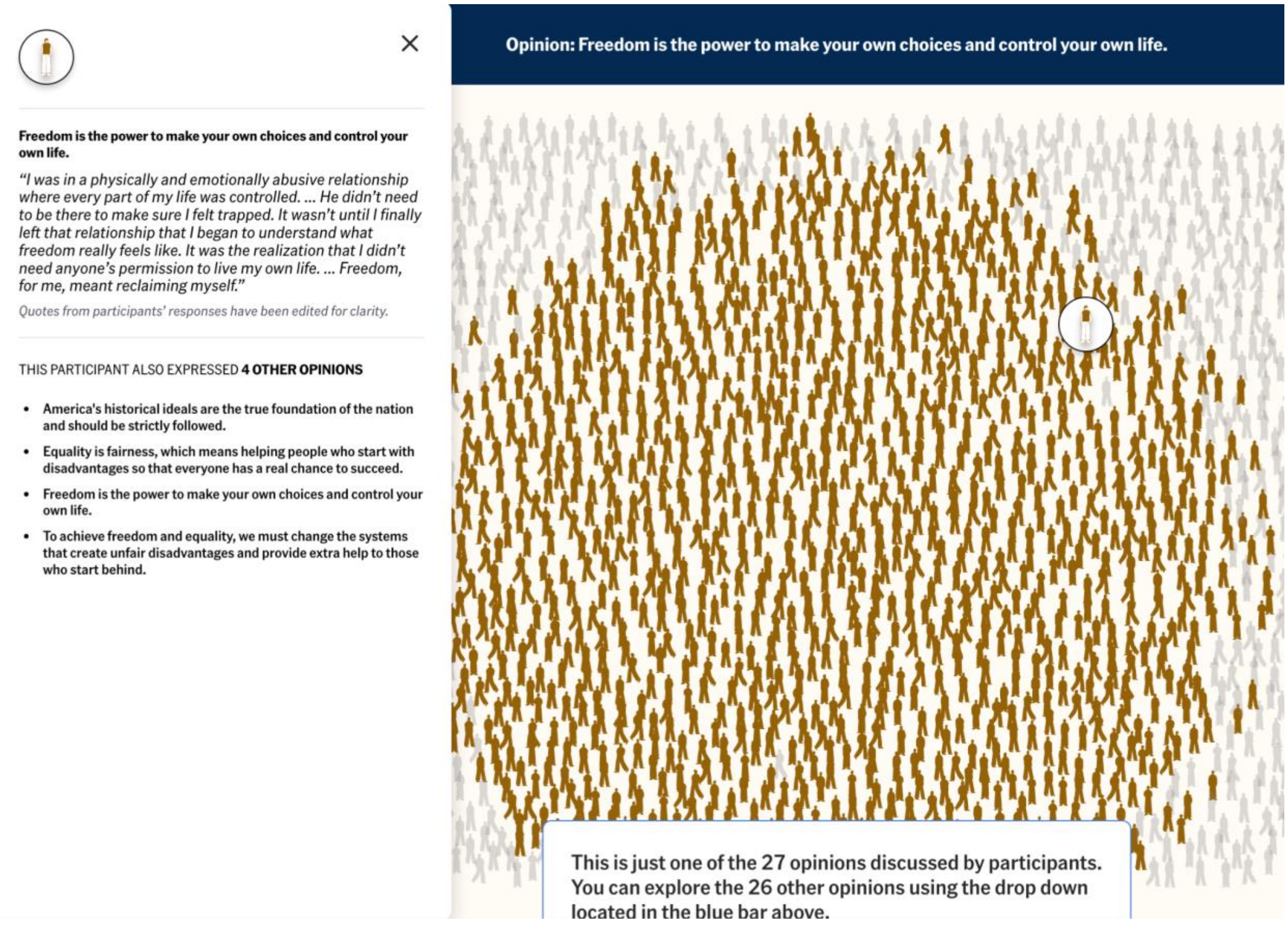

The interactive report ends by inviting viewers in to answer questions and test how their views compared with participants in an interactive graphic, illustrated in Figure 8. The icon in the middle is meant to represent the viewer and the circle of like-minded peers who agree. This circle swells with each answer, visualizing the sense of belonging to a wider community of Americans and the significant common ground we have on core political values.

*Figure 8: We the People Opinion Map B*

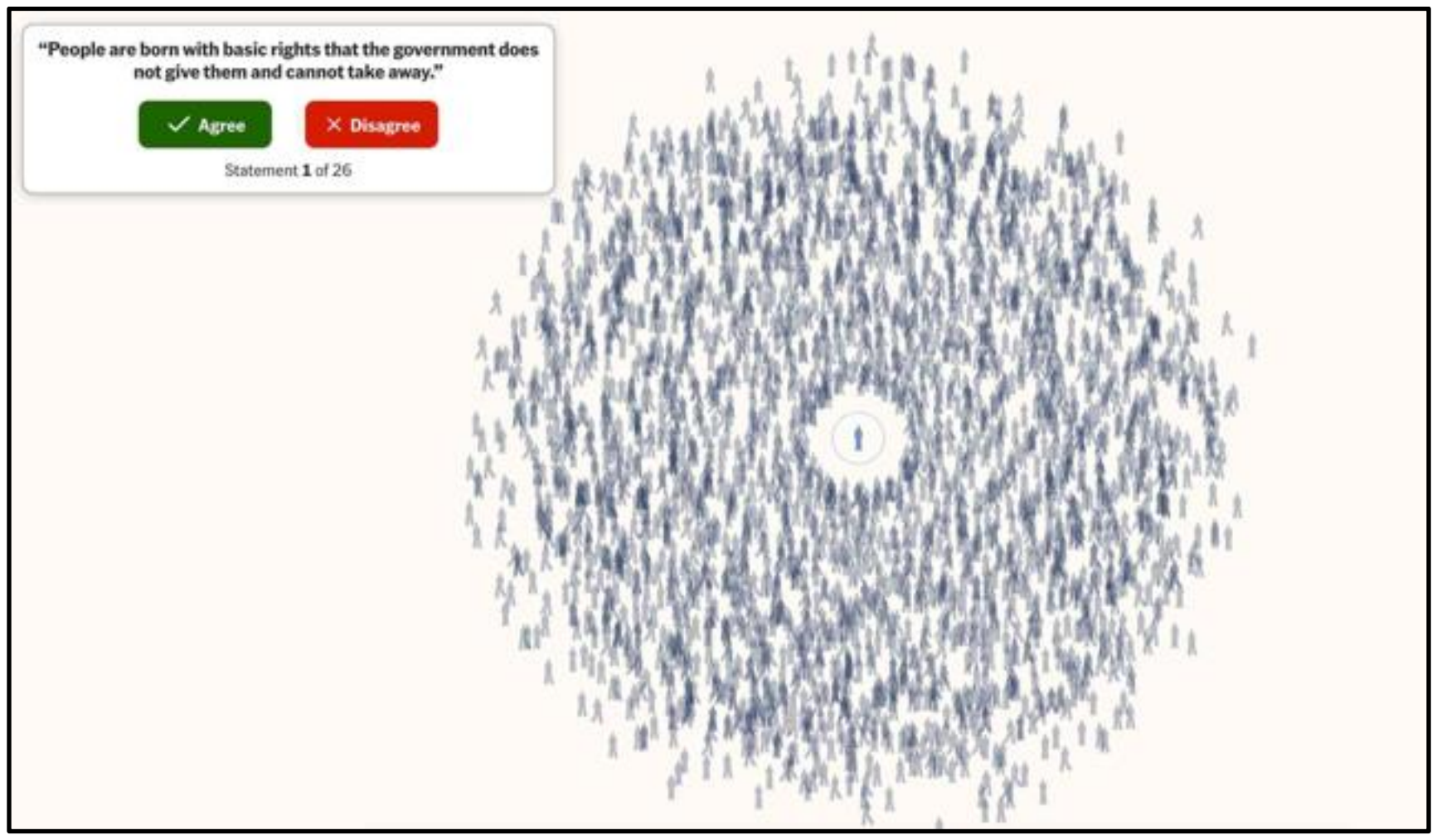

The results offer a powerful alternative to "us vs. them" ways of understanding pluralism in Americans' opinions. The process generated 26 summary statements, most with over 80 percent agreement, revealing substantial common ground often hidden in conventional representations.

While the interactive report is a powerful tool for pluralistic data storytelling, the effects were even more significant for those who participated in the conversation: participants reported high levels of recognition and inter-group understanding. 94 percent felt their views were accurately represented, 89 percent believed the process expanded the range of voices heard, and 97 percent expressed interest in participating again. Despite never meeting in person and engaging asynchronously online for less than an hour each, three out of four participants gained a better understanding of opposing viewpoints.

By visualizing commonalities alongside difference, the project reduces the "perception gap" and shows how AI-enabled, deliberative design can transform not only what people think, but how they see one another—as part of a shared, complex "we."The nationwide conversation surfaced patterns across thousands of voices, including their shared hopes, tensions, contradictions, and points of overlap. The interactive visualization helps the public see themselves included in the larger opinion map of all of America. Not "us versus them," but a dense landscape of perspectives that can be seen, explored, and engaged. This created a "living, breathing portrait" of the nation.

This was a promising pilot; it is important to test conversations and visualizations like this in a variety of contexts and on even more politically salient topics. Jigsaw open sourced the code both to run the three part conversation and visualize the interactive report, making it easier for others to conduct similar such dialogues and share these conversations through pluralistic, interactive visualizations.

### Conclusion

Research suggests that changing the visual and interactive frameworks through which we consume political information can bridge the perception gap and cultivate a more informed, less polarized electorate. The takeaway for journalists, designers, and consumers is clear: data is never neutral, and can even be prosocial. A bar chart showing a 30% gap between the parties tells one story. But a distribution chart showing the 70% of the area where those groups overlap tells a story that emphasizes their commonality and is perhaps more truthful.

Deliberative technologies are new tools to help communities navigate and make sense of different points of view. They provide a safe architecture for listening, and offer affordances for scaling dialogue. Some deliberative technologies, like those outlined here including Pol.is, Consider.it, and Jigsaw's sensemaking technology, offer new ways both to converse across differences and to do pluralistic data storytelling. These visualizations help to enrich oversimplified narratives and clarify misperceptions.

Visual data storytelling impacts polarization. In Jigsaw's research about the national conversation, *We the People*, 75% of participants reported feeling like they had a better understanding of different viewpoints, even if they disagreed with them. We need to make intentional design choices to produce visualizations that highlight where we differ, while also accurately reflecting the common ground that exists. If we change how we look at the data, we might just change how we look at each other.

**Acknowledgements:** We the People data visualization was the hard work of many people but special thanks to Matt Daniels, Peter Wiegand and Thea Mann for their leadership throughout.